\documentclass[journal,twocolumn]{IEEEtran}
\RequirePackage[OT1]{fontenc}
\usepackage{enumitem}
\usepackage{url}
\usepackage{bm}
\usepackage{graphicx}
\usepackage{amsfonts,amsmath}
\usepackage{float}

\usepackage{subcaption}

\usepackage{listings}
\usepackage{pgfplots}

\usepackage{hyperref}

\pgfplotsset{compat=1.17}

\begin{document}
	\title{Hardware Synthesis of State-Space Equations; Application to FPGA Implementation of Shallow and Deep Neural Networks}
	\author{Amir-Hossein~Kiamarzi, Pezhman~Torabi and Reza~Sameni$^*$ \thanks{$^\textnormal{*}$Department of Biomedical Informatics, Emory University, 101 Woodruff Circle, Atlanta, GA 30322, US  (e-mail: \url{rsameni@dbmi.emory.edu})}%
	}
	\maketitle
	
	\begin{abstract}

Nowadays, shallow and deep Neural Networks (NNs) have vast applications including biomedical engineering, image processing, computer vision, and speech recognition. Many researchers have developed hardware accelerators including field-programmable gate   arrays (FPGAs) for implementing high-performance and energy efficient NNs. Apparently, the hardware architecture design process is specific and time-consuming for each NN. Therefore, a systematic way to design, implement and optimize NNs is highly demanded. The paper presents a systematic approach to implement state-space models in register transfer level (RTL), with special interest for NN implementation. The proposed design flow is based on the iterative nature of state-space models and the analogy between state-space formulations and finite-state machines. The method can be used in linear/nonlinear and time-varying/time-invariant systems. It can also be used to implement either intrinsically iterative systems (widely used in various domains such as signal processing, numerical analysis, computer arithmetic, and control engineering), or systems that could be rewritten in equivalent iterative forms. The implementation of recurrent NNs such as long short-term memory (LSTM) NNs, which have intrinsic state-space forms, are another major applications for this framework.

As a case study, it is shown that state-space systems can be used for the systematic implementation and optimization of NNs (as nonlinear and time-varying dynamic systems). An RTL code generating software is also provided online, which simplifies the automatic generation of NNs of arbitrary size.
\end{abstract}
	
	\section{Introduction}
	Deep Neural Networks (DNNs) have become extremely popular in various applications such as computer vision, video analytics, robotics, speech recognition, natural language processing, and web search \cite{hauswald2015sirius,graves2013speech,krizhevsky2012imagenet,hauswald2015djinn,wu2015deep,coates2013deep}. While DNNs are highly advantageous, they require a notable amount of power for computation. Therefore, researchers have focused on specialized accelerators for DNNs \cite{chen2014diannao,chen2014dadiannao,zhang2015optimizing,farabet2011neuflow}, and other workloads \cite{venkatesh2010conservation,govindaraju2011dynamically,gupta2011bundled,putnam2014reconfigurable,liu2015pudiannao,du2015shidiannao,mahajan2016tabla}. Application-specific integrated circuits (ASICs) provide significant performance and efficiency gains for DNNs \cite{chen2016eyeriss,sim201614,qadeer2013convolution,conti2015ultra}, but they may not cope with the ever-evolving DNN models. Furthermore, the high cost of design and lengthy design period of ASICs and customized cores are among the greatest disadvantages of such accelerators. FPGAs are an attractive alternative for accelerating DNNs, as they provide an average efficiency between ASICs and the programmability of general-purpose processors. FPGAs still require extensive hardware design expertise and long design cycles.
	
	Several research studies have been done to provide an ad-hoc FPGA-based accelerator for specific DNN models \cite{gokhale2014240,farabet2010hardware}. However, with a change in DNN models, the designer should change the designed architecture, which is time demanding. Therefore, the lack of a systematic method to design the desired architecture remains an open problem. In addition to DNNs, there are many other applications, which rely on iterative methods with repeated, yet similar calculations. Finite and infinite impulse response filters \cite{analog2003mixed}, Kalman and adaptive filters \cite{meyer2007digital}, discrete Fourier transforms \cite{oppenheim2019discrete} are some examples of signal processing applications of iterative methods that are widely used in digital systems. Other examples include numerical analysis algorithms such as the Bisection, fixed-point \cite{zarowski2004introduction}, and the Newton-Raphson root-finding method \cite{deschamps2006synthesis}, and the Euler method to solve ordinary differential equations (ODEs) \cite{suli2003introduction}.
	
	From the implementation and optimization perspective, iterative algorithms are difficult to parallelize. To highlight the context of the proposed framework, consider a first-order iterative calculation example:
	\begin{equation*}
	y[n] = \alpha y[n-1] + x[n]
	\end{equation*}
	which is the building block in many signal processing and numerical analysis algorithms.	Accordingly, $y[n]$ has one sample dependency to $y[n-1]$, which makes this form of the calculation inappropriate for parallel implementations of the multiplication operator. To resolve this issue, various approaches have been proposed, such as rewriting the iterative equation in equivalent forms (adding excess poles and zeros that cancel from the numerator and denominator of the input-output transfer function), or by using poly-phase calculation techniques \cite{meyer2007digital}. However, these techniques are highly ad hoc and may not be systematically extended to other iterative algorithms. Therefore, iterative calculations are commonly a bottleneck for parallel implementations.

	In this work, we develop a high-level synthesis approach for the hardware implementation of iterative algorithms based on a \textit{state-space} formulation. It is shown that the proposed design flow can be also used for systems that are not originally iterative, but could be rewritten and implemented in an iterative state-space formulation. DNNs are one of this kind. While, many ad-hoc methods have been suggested in the literature to implement these algorithms on FPGAs, the purpose of this work is to propose a systematic workflow, which is applicable to a broad class of algorithms, which can be implemented in state-space form.
	
	The paper is organized as follows: Section~\ref{sec:background} reviews the state-space model, the finite state machine, and the similarity between them. In Section~\ref{sec:proposedmethod}, the proposed state-space based workflow is introduced. The application of the proposed workflow for implementing an arbitrary neural network (scalable to DNNs) is presented in Section~\ref{sec:case_study}, followed by concluding remarks and future perspectives.
	
	\section{Background}
	\label{sec:background}
	The basis of the proposed workflow are state-space representations, finite-state machines, and their analogies.
	\subsection{State-space model}
\begin{figure}[tb]
		\centering
		\includegraphics[width=0.95\columnwidth]{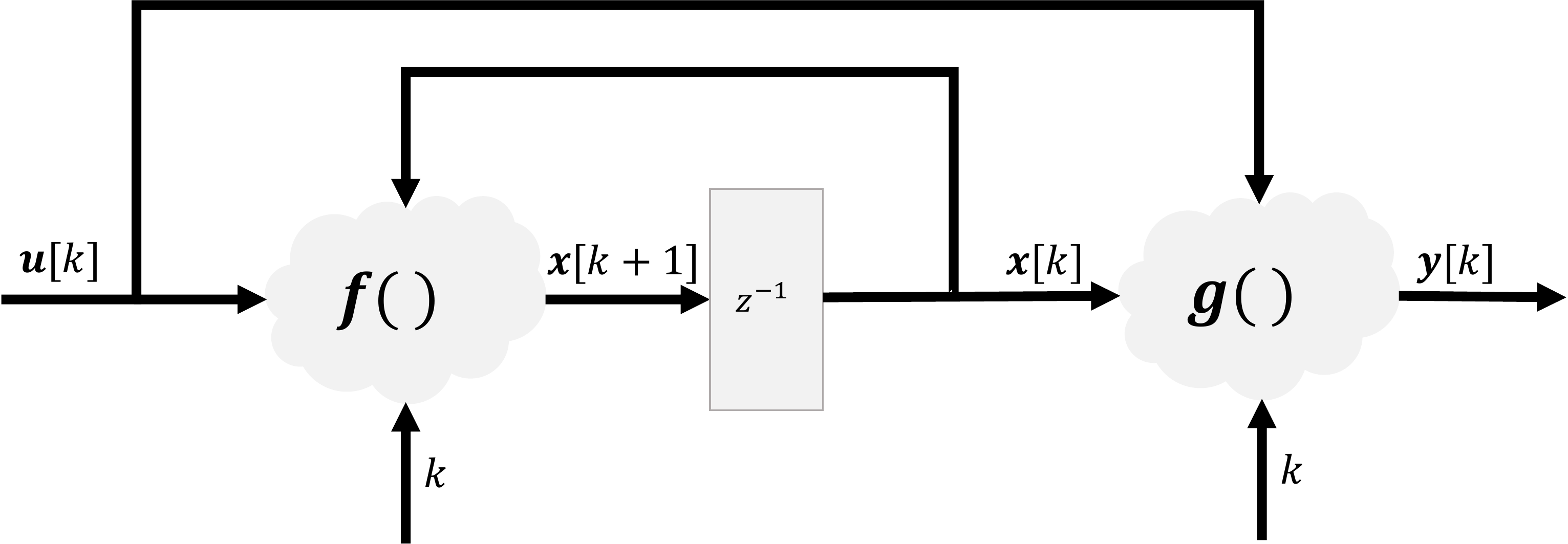}
		\caption{Block diagram of a discrete-time dynamic system, adapted from \cite{ogata1995discrete}} 
		\label{fig:dynamicsystem}
	\end{figure}

	The state-space is the description of a system's dynamics in terms of first-order difference (or differential) equations, which can be combined into a set of first-order vector/matrix difference equations. The use of the vector-matrix notation simplifies the mathematical representation of the system and permits the development of generic mathematical frameworks to study systems of diverse origin. General properties such as stability, parameter identifiability, controlability of the states, state observability, state estimation and control are among the numerous physical/mathematical properties of dynamic systems, which can be studied by using the state-space representation \cite{kailath1980linear,ogata1995discrete}. The state-space form of a generally time-varying (linear or nonlinear) discrete-time dynamic system is as follows (cf. Fig.~\ref{fig:dynamicsystem}):
	\begin{equation}
	\begin{array}{rl}
	\mathbf{x}[k+1] = \mathbf{f}\left(\mathbf{x}[k],\mathbf{u}[k],k\right) \\
	\mathbf{y}[k] = \mathbf{g}\left(\mathbf{x}[k],\mathbf{u}[k],k\right) 
	\end{array}
	\label{eq:statespace}
	\end{equation}
	where $\mathbf{u}[k]$, $\mathbf{x}[k]$ and $\mathbf{y}[k]$ are input, state, and output vectors of the system.
	
	\subsection{Finite-state machine}
\begin{figure}[tb]
			\centering
			\includegraphics[width=0.95\columnwidth]{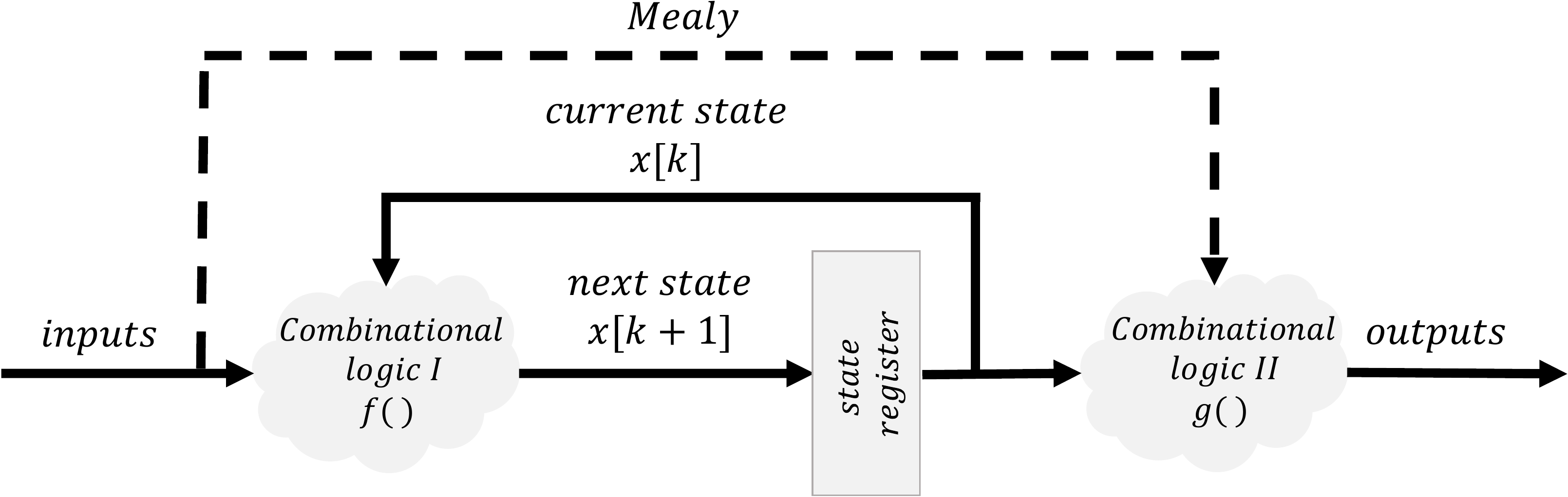}
			\caption{Composition of a Mealy/Moore machine implementation of an FSM \cite{brown2007fundamentals}.}
			\label{fig:MealyMoore}
		\end{figure}

	Finite-state machines (FSM) are analogous with state-space models in system theory. In synchronous digital design, the results of operations performed on a sequence of data are stored in registers and memories according to control signals generated by FSMs. An FSM is a sequential digital circuit with an $N$-bit state register to store the current state of the operations. The hardware implementation of an FSM is composed of a \textit{combinational logic} and \textit{sequential logic}. The combination logic computes the output and the next state based on the current state and input; whereas, the sequential part has a state register to preserve previous states, as required for the implementation of logic circuits with memory. FSMs are implemented in two forms, known as the Mealy and Moore state machines. In a Mealy machine, the output is a function of the current state and the input, whereas in a Moore machine, the outputs are only a function of the current state, as shown in Fig.~\ref{fig:MealyMoore}.
	
	
	Mathematically, an FSM can be formulated as a sextuple $\left(U, Y, X, \mathbf{x}[0], \mathbf{f}(\cdot), \mathbf{g}(\cdot) \right)$, where $U$ and $Y$ are sets of inputs and outputs, X is a set of states, $\mathbf{x}[0]$ is the initial state, and $\mathbf{f}(\cdot)$ and $\mathbf{g}(\cdot)$ are the next state and output functions, respectively \cite{khan2011digital}. The bold notation represents the fact that the inputs, states, and outputs can generally be vectors (arrays) of binary values. The expression for next state computation can be written as:
	\begin{equation}
	\mathbf{x}[k+1] =
	\mathbf{f}\left(\mathbf{x}[k], \mathbf{u}[k]\right)
	\label{eq:FSMNextState}
	\end{equation}
	where $\mathbf{x}[k] \in X$, $\mathbf{u}[k] \in U$, and index $k$ is the time index of the current input and state vectors, and $\mathbf{x}[k+1]$ is the next state. For Mealy and Moore machines, the expressions for computing output can be written as:
	\begin{equation}
	\begin{split}
	&\mbox{Mealy machine: } \mathbf{y}[k] =
	\mathbf{g}\left(\mathbf{x}[k],\mathbf{u}[k]\right)\\
	&\mbox{Moore machine: } \mathbf{y}[k] =
	\mathbf{g}\left(\mathbf{x}[k] \right)
	\end{split}
	\label{eq:FSMOutputEq}
	\end{equation}
	which shows that the only difference between the two machines, is that the output equation of a Moore machine is not a function of the current input, i.e. the input only appears in the output through the state variables.
	\subsection{The analogy between FSM and state-space models}
	\label{sec:FSMvsSSanalogy}
	From the above review, the analogy between state-space models and FSMs is evident. Considering the combinational logics in Fig.~\ref{fig:MealyMoore} as the state-update and output-update functions in Fig.~\ref{fig:dynamicsystem}, and modeling a shift-register with the z-transform delay element $z^{-1}$, FSMs are principally a \textit{finite-field arithmetic} special case of general state-space models. Accordingly, the only difference between the two entities is that in an FSM, the inputs, states, and outputs are selected from a finite set, and the combinational and sequential arithmetics are performed on these finite sets, while in real-valued state-space models, the quantities are real-valued. Nevertheless, in high-precision numeric calculations, where the FSM variables represent sequential numbers in fixed- or floating-point format in high number of quantization bits, the difference between an FSM and state-space models seemingly vanishes, since the effect of finite-field arithmetic quantizations becomes negligible and may be modeled as \textit{process noise} and \textit{output noise}, as commonly used in classical state-space modeling and estimation schemes.
	
	The above analogy enables hardware architects to benefit from the rich and profound mathematical rigor of state-space models, in hardware design and analysis. The applications include fault tolerant hardware design (both during fabrication and operation), metastability analysis of logic circuits, pseudo-random number generator design and analysis, limit cycles etc. Herein, we focus on the implementation advantages of state-space forms, which enables a systematic approach for pipelining iterative calculations.
	
\begin{center}
		\begin{figure}[tb]
			\begin{subfigure}{\linewidth}
				\centering
				\includegraphics[width=0.7\linewidth]{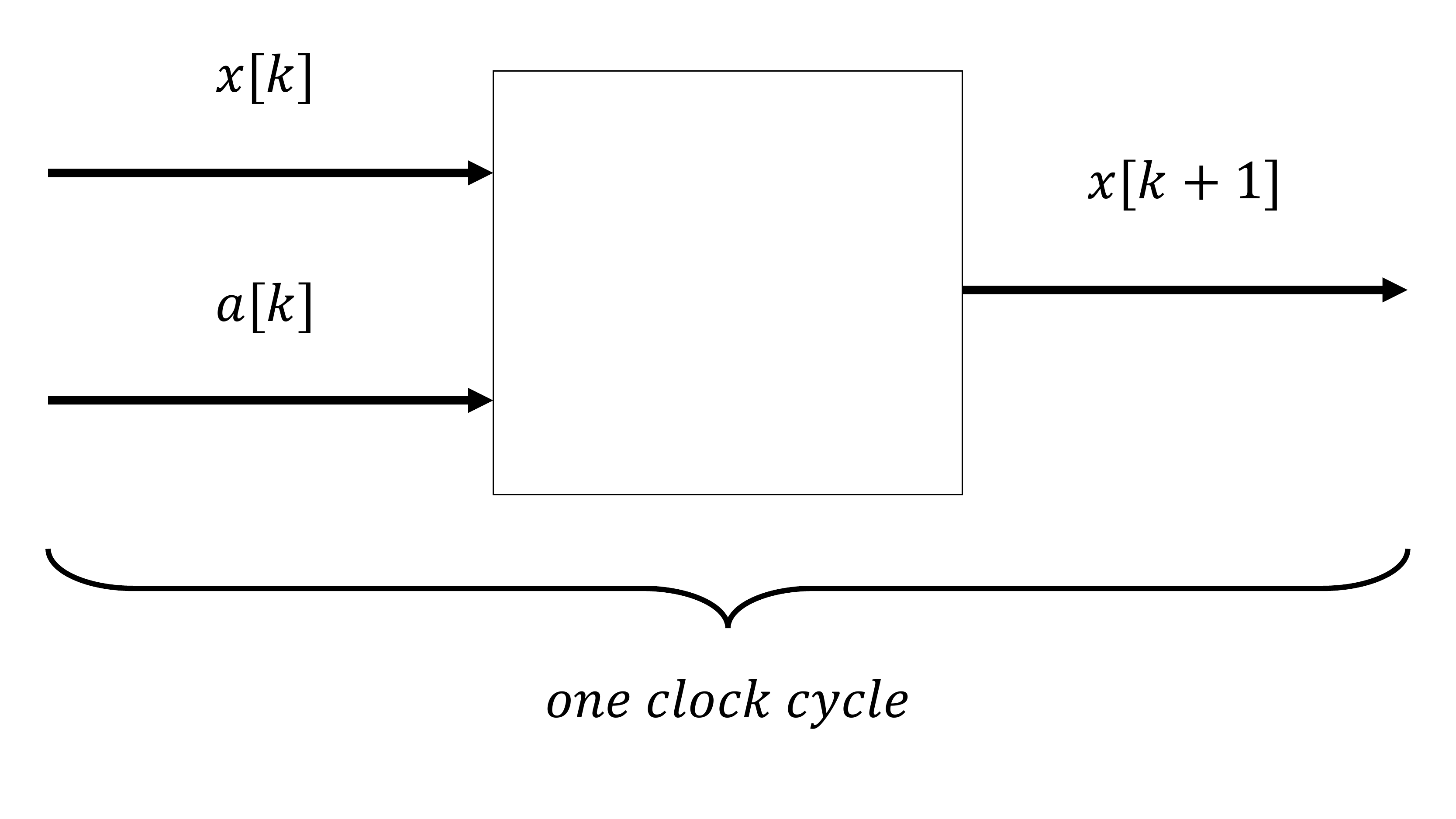}
				\caption{}
			\end{subfigure}
			\begin{subfigure}{\linewidth}
				\centering
				\includegraphics[width=0.7\linewidth]{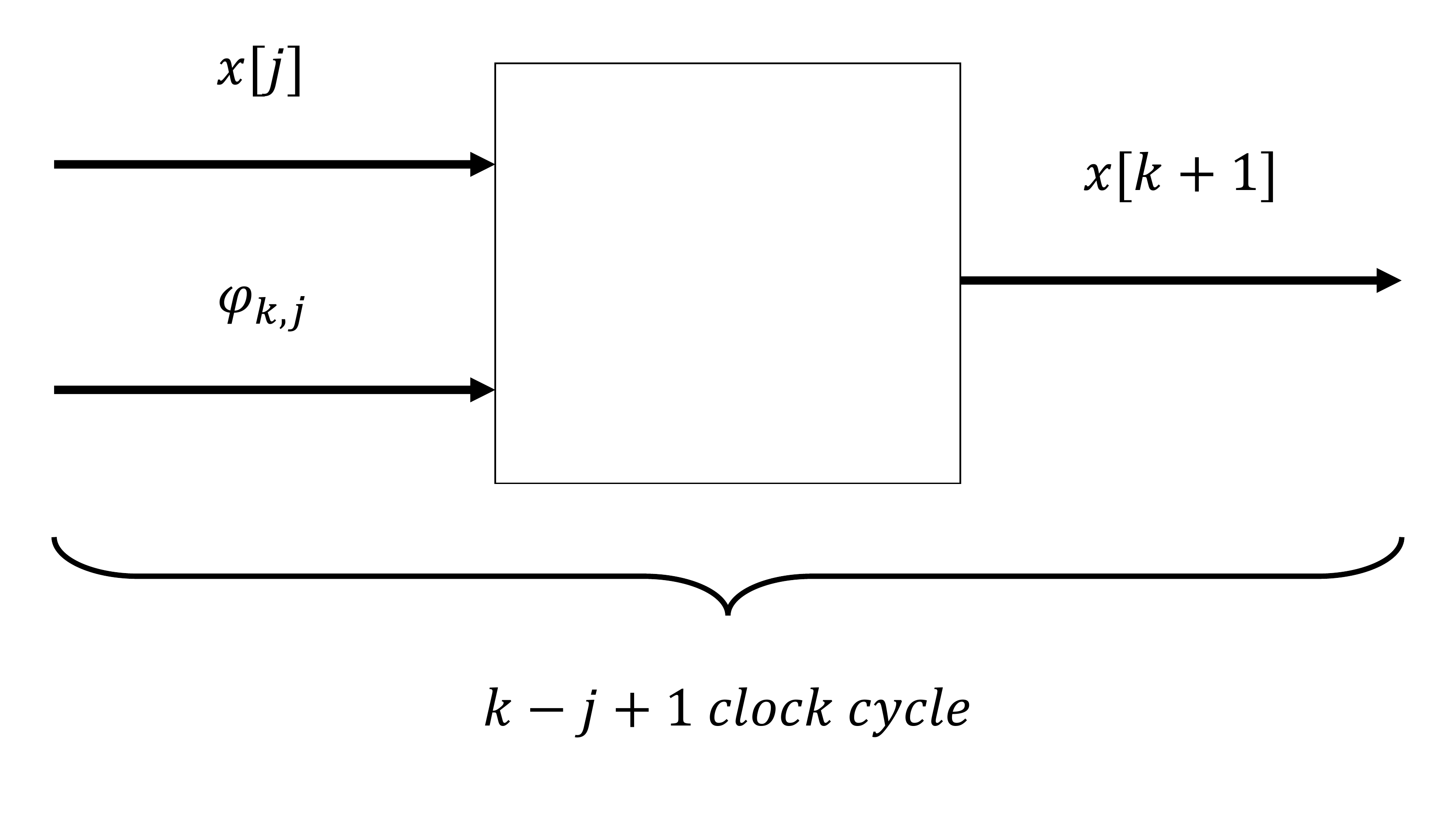}
				\caption{}
			\end{subfigure}
			\begin{subfigure}{\linewidth}
				\centering
				\includegraphics[width=.95\linewidth]{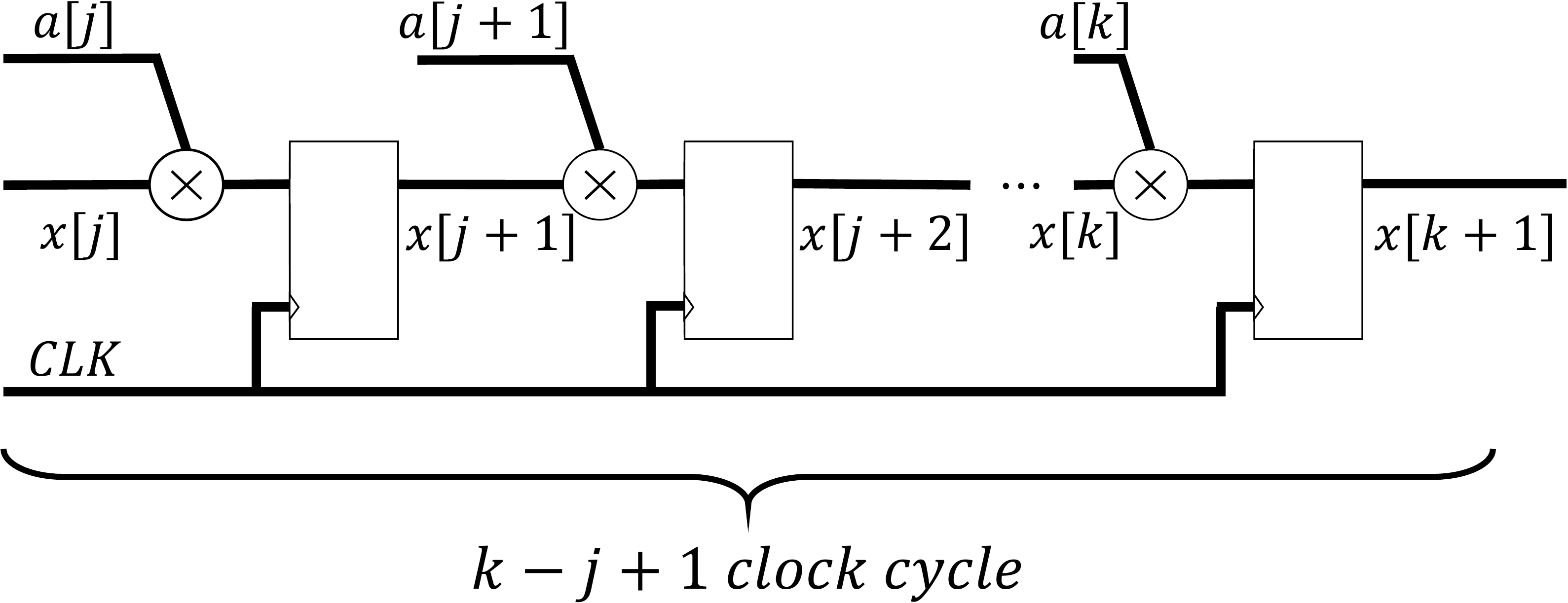}
				\caption{}
			\end{subfigure}
			\caption{The state-space model (\ref{eq:state_space_linear}) before and after using the $j-step$ state transition matrix.}
			\label{fig:state_transition_matrix}
		\end{figure}
	\end{center}

	Let us consider a special case of state-space models, where the system's dynamics is linear: 
	\begin{equation}
	\mathbf{x}[k+1] = \mathbf{A}[k] \mathbf{x}[k]
	\label{eq:state_space_linear}
	\end{equation}
	The above can be extended to a $j$-step update:
	\begin{equation}
	\mathbf{x}[k + 1] = 
	\bm{\Phi}_{kj}
	\mathbf{x}[k-j]
	\label{eq:state_j_step_update}
	\end{equation}
	where $\bm{\Phi}_{k,j}=\mathbf{A}[k]
	\mathbf{A}[k-1]\ldots \mathbf{A}[k-j]$ is known as the \textit{state transition matrix}. Although (\ref{eq:state_space_linear}) and (\ref{eq:state_j_step_update}) are identical, from the computational perspective (\ref{eq:state_j_step_update}) can be more advantageous, since the next state vector depends on the state vector from $j$ steps before, rather than the immediate previous state. Therefore, the required state-update calculation can be pipelined by using the $j$-step state-transition matrix $\bm{\Phi}_{kj}$ in (\ref{eq:state_j_step_update}), instead of $\mathbf{A}[k]$, as illustrated in Fig. \ref{fig:state_transition_matrix}.
	
\begin{figure}[tb]
		\centering
		\includegraphics[trim=1in 1.0in 2in 1.5in,clip,width=\columnwidth]{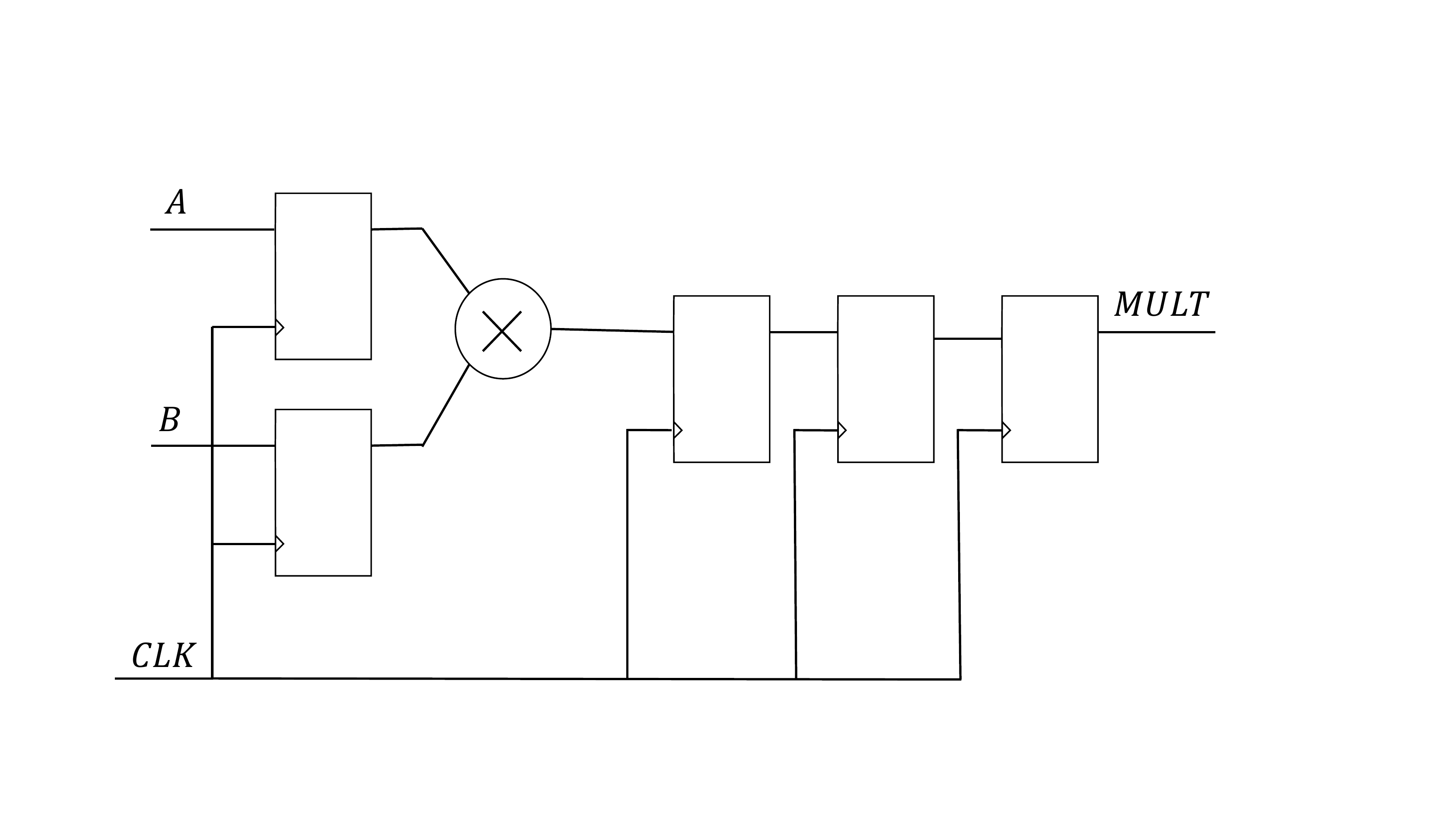}
		\caption{Block diagram of the XILINX multiplier \cite{xilinxXSTuserguide}, which is automatically replaceable by a resource/clock optimized equivalent pipelined multiplier by XST}
		\label{fig:XILINX_mult_pipe}
	\end{figure}

	This pipelining technique, which is mathematically based on the concept of state transition matrices/functions in state-space models, has been utilized in some hardware synthesis tools for the automatic pipelining of hardware calculations. For example, in the XILINX XST synthesis tool ``in order to increase the speed of designs with large multipliers, XST can infer pipelined multipliers. By interspersing registers between the stages of large multipliers, pipelining can significantly increase the overall frequency of the design  \cite{xilinxXSTuserguide}.'' Accordingly, when register transfer level (RTL) code implicitly, or explicitly defines a multiplier as in Fig.~\ref{fig:XILINX_mult_pipe}, without any external dependencies to the inter-register values of this multiplier, XST automatically replaces the multiplier, with an efficient pipelined implementation, which has the same number of flip-flops from its input to outputs, but is more efficient in terms of either resources or its maximum throughput (depending on the optimization criterion). This is of course a very limited utilization of the vast advantages of state-space representations. In the sequel, we propose a general workflow, which enables similar optimizations, for a broad class of iterative calculation schemes.
	
	\section{Proposed workflow}
	\label{sec:proposedmethod}
\begin{figure*}[tb]
		\centering
		\includegraphics[width=0.98\textwidth]{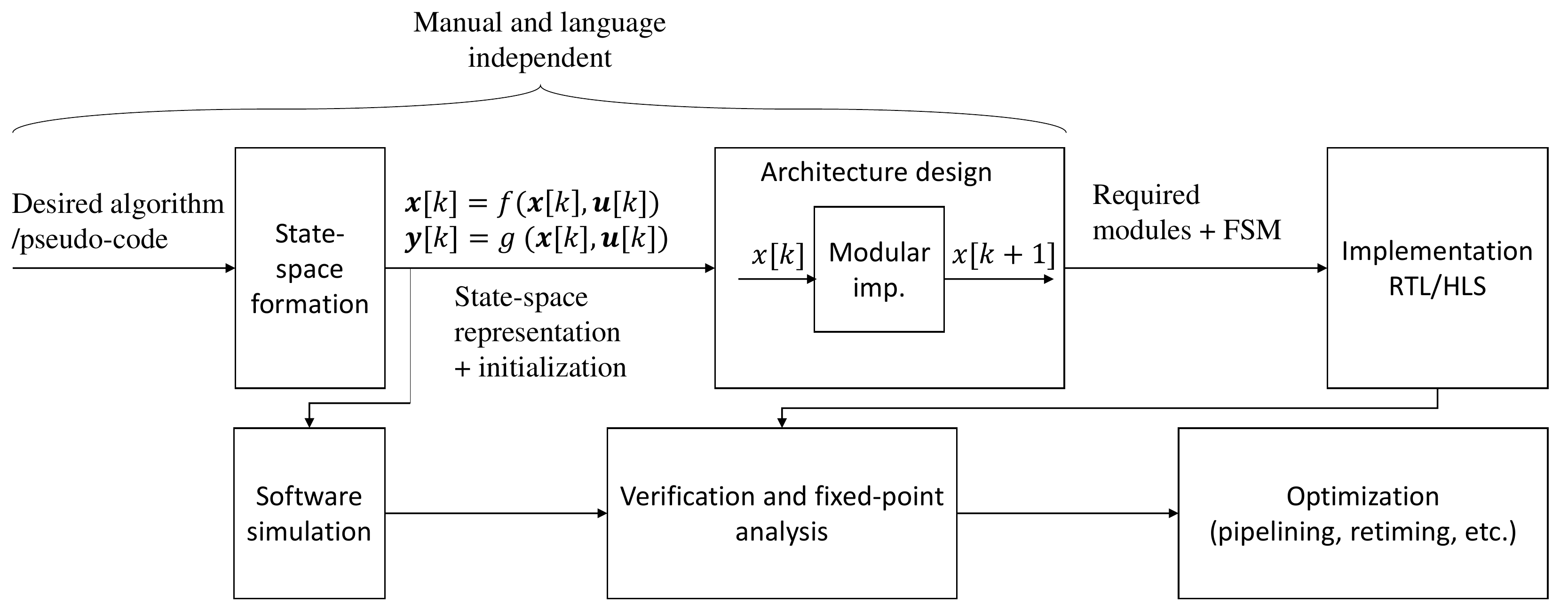}
		\caption{An overview of the proposed workflow.}
		\label{fig:proposed_method_diagram}
	\end{figure*}

	The proposed workflow consists of a systematic approach to convert iterative algorithms to a synthesizable and optimal hardware implementation. The required stages as shown in Fig.~\ref{fig:proposed_method_diagram} are detailed in the sequel.
	\subsection{State-space formation}
	The design entry is assumed to be in pseudo-code, hardware description language, or even as a software implementation. At the first stage of the workflow, the design entry should be converted to a state-space representation. The state-space form of a dynamic system is not unique. In system theory, the state-space form which satisfy the implementation requirements are used. For example, some of the important properties of a dynamic representation include observability, controllability, reachability, and constructibility \cite{kailath1980linear}, which depending on the dynamics of the system (linear or nonlinear), can be achieved by using specific forms known as canonical forms \cite{kailath1980linear}. In control theory, there are systematic ways for obtaining these canonical forms by methods such as \textit{nested programming} or \textit{partial-fraction-expansion} \cite{ogata1995discrete}.
	
	\subsection{Software simulation}
	Having the state-space form of the system's equations in the form of (\ref{eq:statespace}), plus the initial values of the state variables, one can recursively calculate the output of the system (in most cases numerically), and to verify the state-space representation versus the expected input-output relationship of the original system under different test cases (combinations of input-output vectors and parameter sets). This step, which is carried out either manually or in a software environment, verifies the correctness of the state-space representation, prior to its hardware implementation. MATLAB Simulink, SyestemC, and Intel's Data Parallel C++ (DPC++) language and similar tools provide strong simulation and debugging environments, which can be used during the verification stage.
	
	\subsection{Fixed-point analysis}
	Although dedicated floating-point units (FPUs) are becoming more and more popular on FPGA devices, the majority of high-performance FPGA-based computations are still based on fixed-point and mixed-precision (combinations of fixed-point representations with different word and fractional lengths at different points of the system) implementations. Therefore, fixed-point analysis is a required step before hardware implementation. At this step, the number of required bits for representing the state-space parameters, inputs, states and outputs are selected by analyzing the accumulative effects of quantization error in the output vs the target signal-to-noise ratio (SNR). In this context, a major advantage of state-space forms is that one can systematically analyze the effect of quantization noise. In fact, the quantization noise can be modeled as random noises that are added to the inputs, state variables and the output, and one can specifically calculate (analytically or numerically) the stability of the dynamic model, with respect to quantization errors at variable numbers of quantization bits.

	\subsection{Architecture design}
	Since the system's equations have been converted into state-space form, the architecture consists of three clear elements: 1) the sequential logic implemented by a sequence of flip-flops that connect previous states to next states, 2) the combinational logic, which is the linear or nonlinear transforms corresponding to $\mathbf{f}(\cdot)$ and $\mathbf{g}(\cdot)$ in (\ref{eq:statespace}), and 3) a timing controller for managing the timing between the inputs and outputs, which is generally implemented as an FSM-based state controller. Items 1 and 2 comprise the data path, while item 3 is the control path. Apparently, the structure of the controller depends on whether the data path modules operate at the same clock frequency of the system clock or not. Depending on the case, the controller manages the time/resource sharing between the elements of the data path. 
	
	The combinational modules required to implement a finite-state machine are designed based on the state and output equations of the dynamic system. In (\ref{eq:statespace}), the functions $\mathbf{f}(\cdot)$ and $\mathbf{g}(\cdot)$ can generally be nonlinear. There is a vast literature on the hardware implementation of specific nonlinear transforms. For the non-popular ones, one may use generic Taylor expansion techniques \cite{kilts2007advanced}, CORDIC machines \cite{muller2016elementary}, or look-up table (LUT) based approaches with or without input/output interpolation \cite{parhami2000computer}. 
	\subsection{Implementation}
	After designing the required modules based on the state and output equations, the designed modules are implemented. The modules can be described in arbitrary hardware description languages, such as Verilog and VHDL, or by high-level synthesis tools, such as XILINX Vivado HLS. In the later presented case study, Verilog is used for RTL implementation and Vivado HLS tools are used for implementation on XILINX FPGAs. 
	
	\subsection{Optimization}
	\label{sec:state_space_optimization}
\begin{center}
		\begin{figure}[tb]
			\begin{subfigure}{\linewidth}
				\centering
				\includegraphics[width=0.95\columnwidth]{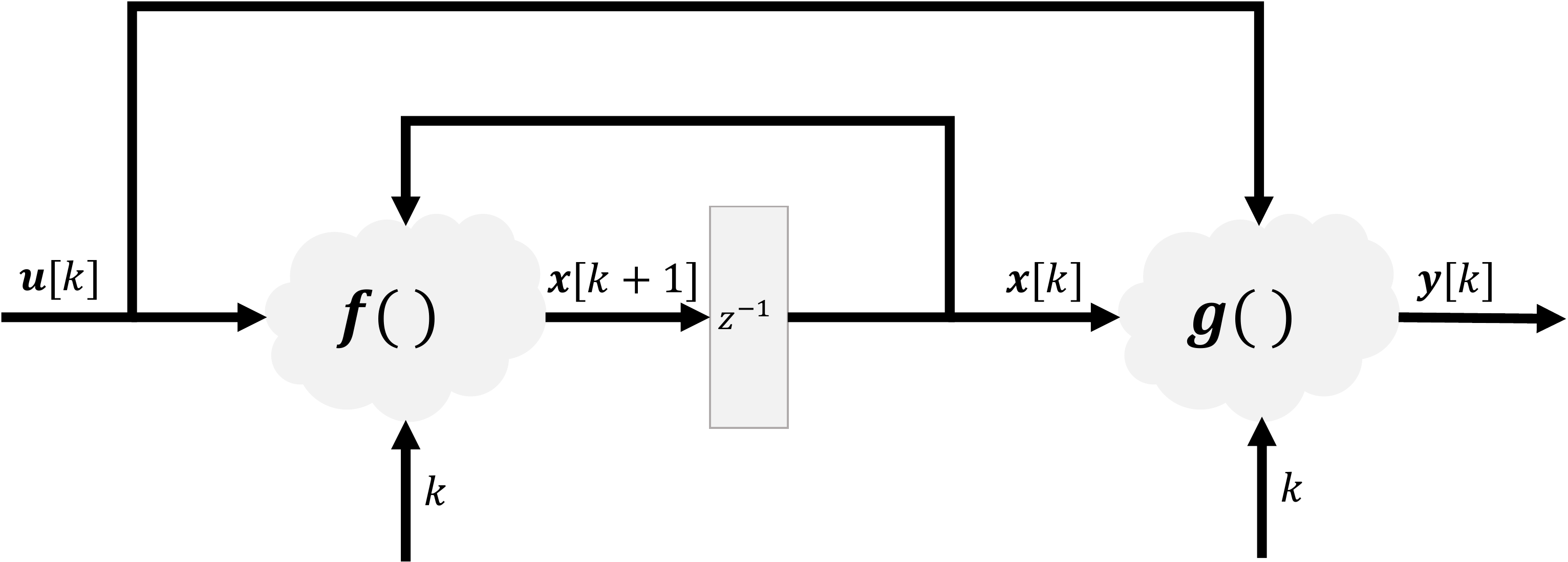}
				\caption{}
			\end{subfigure}
			\begin{subfigure}{\linewidth}
				\centering
				\includegraphics[width=0.95\columnwidth]{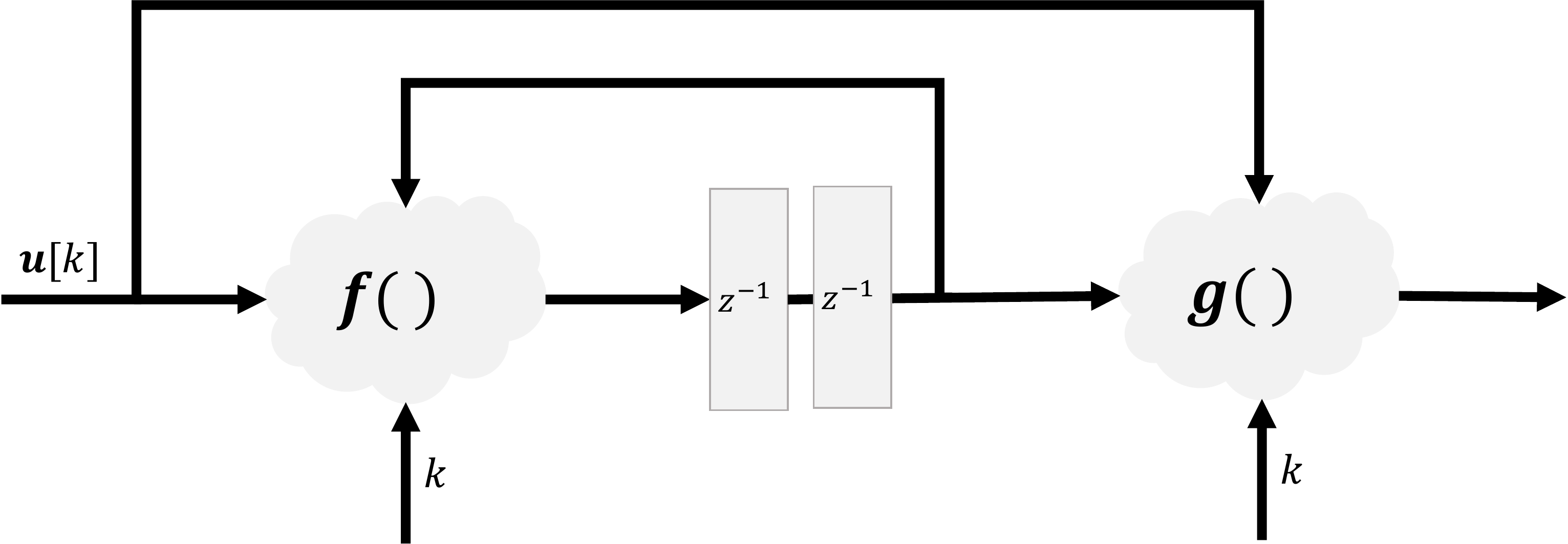}
				\caption{}
			\end{subfigure}
			\begin{subfigure}{\linewidth}
				\centering
				\includegraphics[width=0.95\columnwidth]{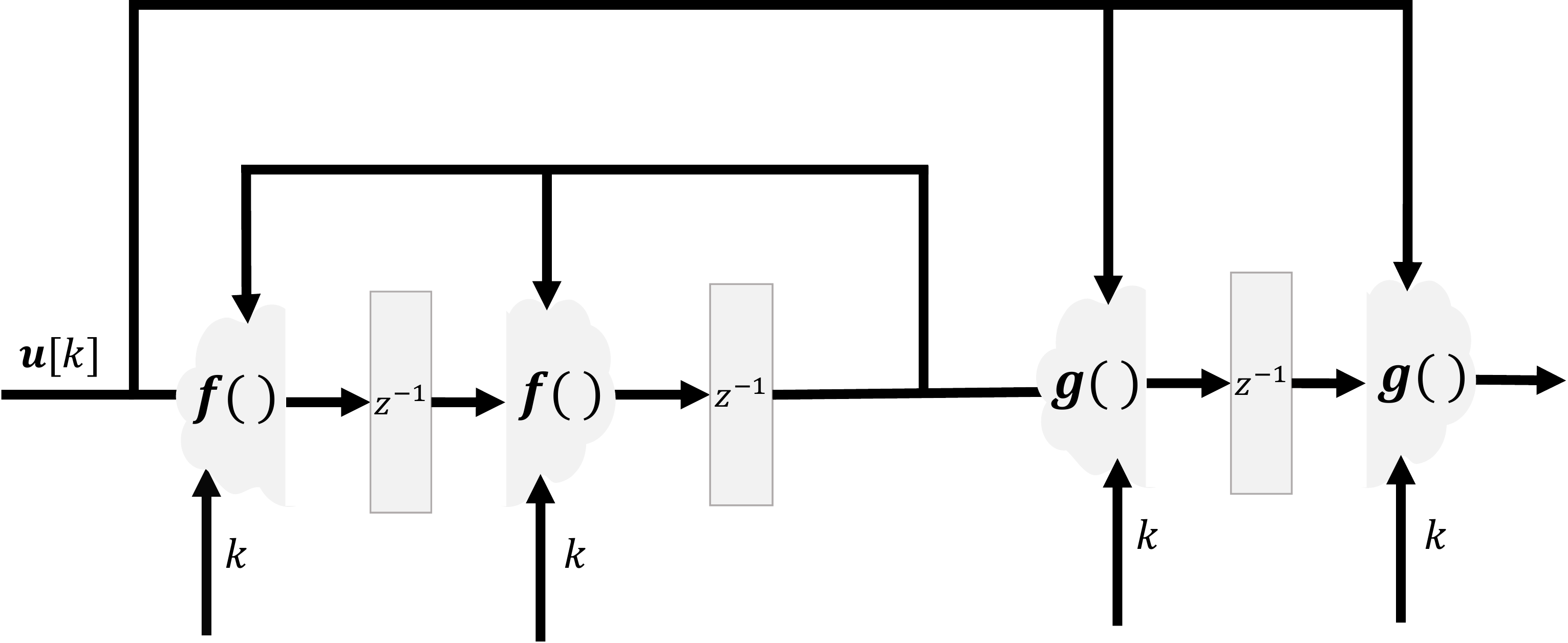}
				\caption{}
			\end{subfigure}
			\caption{The C-slow retiming procedure ($C = 2$)}
			\label{fig:c-slow}
		\end{figure}
	\end{center}

	Up to this point, a synthesizable hardware realization of the state-space equations is available, which is functionally identical to the software-based implementation. However, the design is not necessarily optimal in terms of clock speed or resources. Therefore, after validating the functionality of the hardware implementation, generic optimization techniques such as pipelining, retiming, and C-slow retiming are applied to achieve the desired design requirements. The clear distinction of the combinational and sequential parts in the state-space implementation, plus the state-transition transform detailed in Section~\ref{sec:FSMvsSSanalogy} facilitate the optimization process. To show the optimization procedure, the 2-slow retiming procedure of a state-space circuit representation is shown in Fig.~\ref{fig:c-slow}. 
	
	\section{Case Study: Hardware Implementation of Neural Networks}
	\label{sec:case_study}	
	Recent research in machine learning and pattern recognition applications such as image, speech, and video recognition have shown a significant outperformance of deep neural network (DNN) architectures over traditional algorithms based on handcrafted features and models. Neural networks (NNs) of different architectures have turned into standard building blocks, which are connected hierarchically to perform tasks such as denoising, classification, clustering, regression, etc. This ability, together with the high amount of data required for training DNNs has motivated the design of hardware accelerators for DNN applications \cite{DBLP:journals/corr/abs-1712-08934}.
	
\begin{figure*}[tb]
		\centering\includegraphics[width=0.9\textwidth]{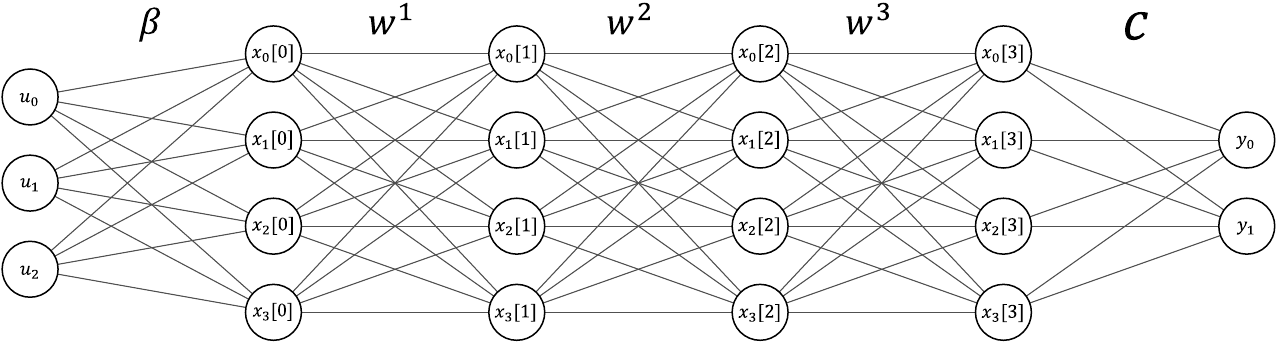}
		\caption{The implemented multi-layer neural network architecture}
		\label{fig:NNarchitecture}
	\end{figure*}

	Due to this popularity, we have adopted a NN architecture as case study. The case study is a good example for illustrating the concept of state-space based implementation of iterative nonlinear algorithms. In the sequel, we detail the stages of the proposed workflow for implementing a rather generic NN architecture on FPGA. As shown in Fig.~\ref{fig:NNarchitecture}, the NN is a multi-layer perceptron composed of three inputs, two outputs, four fully connected hidden layers, and $\tanh(\cdot)$ activation functions. Due to the scalability of the architecture, the implementation is easily and systematically extendable to DNNs of variable dimensions (limited by the available resources on the selected FPGA).
	
	\subsection{State-space formation}
	NNs are not intrinsically iterative, but depending on the architecture of interest and the data flow rate into the NN, they can be written in iterative form. Consider two extreme cases for a NN architecture design:
	\begin{enumerate}
		\item A single node that applies the nonlinear node function on the weighted sum of the previous nodes. In this case, the same node architecture is shared for the calculations of all nodes and the intermediate inputs/results are read/written from/to a memory, which is managed by a state controller. This architecture results in the least area utilization, but also has the minimum throughput, since the same hardware unit is resource-shared for all calculations.
		\item Each node of the NN is implemented independently and runs in parallel. This architecture results requires the maximum area (no resources shared), but has the advantage of the maximum throughput.
	\end{enumerate}
	Herein, an intermediate approach between the above two cases is considered: we assume that a single layer of the NN is implemented with all nodes functioning in parallel, but the same hardware resources are shared between the different layers. Note however that our code-generator tools (detailed in Section~\ref{sec:hdl_code_generator}) are more generic and are designed to implement HDL codes corresponding to different resource-speed compromising schemes defined by the user.
	
	As noted before, state-space forms are not unique. Due to the layer-wise resource sharing architecture that we have assumed for the NN architecture, we define the outputs of the $k$th layer of the NN as the state-vector $\mathbf{x}[k] = [x_1[k], x_2[k], \ldots, x_M[k]]^T$ ($k = 1, \ldots, N$), $\mathbf{W}[k] = [w_{ij}[k]]$ as the NN hidden-layers' weight matrix, $\mathbf{b} = [b_1, b_2, \ldots, b_M]$ as the biases, $\bm{\beta}[k] = [\beta_{ij}[k]]$ as input weight matrix, $\mathbf{C}[k] = [c_{ij}[k]]$ as output weight matrix, $\mathbf{f}(\cdot)$ as the node function, and $\mathbf{y} = [y_1, y_2, \ldots, y_P]^T$ as the outputs. With these definitions, the NN equations can be written in the following state-space form:
	\begin{equation}
	x_i[k + 1] = 
	f_i \left(
	\sum_{j = 0}^{M} w^k_{ji}x_j[k] + b^i\right) + 
	\sum_{j=0}^{L} \beta_{ji}u_j\delta[k]
	\label{eq:nn_state_update0}
	\end{equation}
	
	\begin{equation}
	y_i = 
	\sum_{j=0}^{P} c_{ij}x_j\lbrack N \rbrack
	\label{eq:nn_output_eqn0}
	\end{equation}
	Or in matrix form:
	\begin{equation}
	\begin{array}{rl}
	\mathbf{x}[k + 1] & =
	\mathbf{f} (\mathbf{W}[k]\mathbf{x}[k]) + \mathbf{b}[k]) + \bm{\beta}\mathbf{u}\delta[k]  \\
	\mathbf{y} & =
	\mathbf{C} \mathbf{x}[N] 
	\end{array}
	\label{eq:nn_state_update}
	\end{equation}
	
	Fig. \ref{fig:NNarchitecture} shows that the implemented neural network, with three input neurons ($L = 3$) and four hidden layers ($N = 4$), each containing four neurons ($M = 4$) and two output neurons ($P = 2$). Note that in this formulation $k$ is used as the NN layer index. Therefore, the typical state-update of a classical state-space equation is equivalent to the propagation of the results across the NN from one layer to another.
	
	\subsection{Architecture design}
\begin{figure}[tb]
		\centering
		\includegraphics[width=0.9\columnwidth]{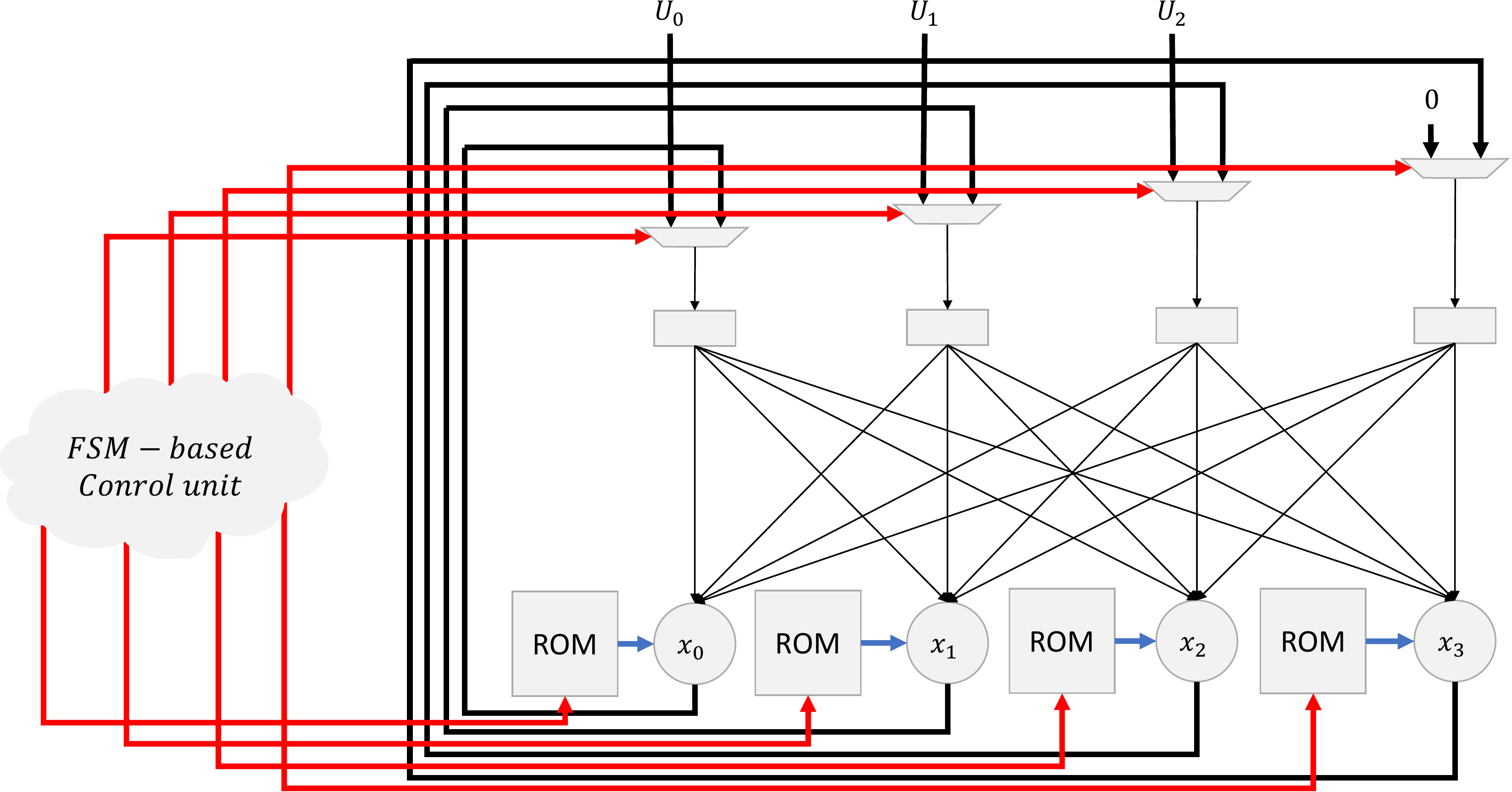}
		\caption{Block diagram of the implemented neural network.}
		\label{fig:NNBlockDiagram}
	\end{figure}

	The base NN architecture consists of an implementation of one layer of a NN consisting of the NN node calculation units, memories for storing inter-layer results, and a state-controller that reads data from the input and controls the data timing and the state evolution of the state-space model, as shown in Fig.~\ref{fig:NNBlockDiagram}.

	In NNs, the nonlinearity of the activity function allows it to learn and model complex tasks from high-dimensional, unstructured, non-linear, and large datasets \cite{tsihrintzis2020machine}. In some cases, the activation function is also used to decide on whether or not the neurons in the next layer should be activated (fired). There are many common nonlinear activation functions, such as \textit{sigmoid}, \textit{tanh}, \textit{softmax}, and \textit{relu}. In this cases study, the $\tanh(\cdot)$ activation function is used as a typical and popular choice, which has a smooth gradient \cite{tsihrintzis2020machine}. Various approaches can be used to implement the $\tanh(\cdot)$ function. For this study, we used a Look-Up Table (LUT) implementation. During the implementation, the $\tanh(\cdot)$ values were calculated and quantized offline in MATLAB, and the samples were stored in a ROM and loaded into the FPGA memories during synthesis.
	
	A significant part of NN computations require multiply-accumulate (MACC) operations. Most contemporary FPGAs have built-in units for MACC operations, such as the DSP48E1 on XILINX FPGAs \cite{7SeriesDSP48E1Slice}.
	
	\subsection{Software simulation}
	Both direct and iterative equations of the introduced NN are simulated and the result is checked to ensure the correctness of the iterative equations. These results are used to perform fixed-point analysis between successive layers of the NN. It is known that in successive calculations (such as the state-space propagations), quantization noise accumulates from the input to the output. In a fully parallel implementation, it is common to use variable numbers of quantization bits across the input-output pipeline. However, for a state-space implementation where the same architecture is resource-shared during state update, we have fixed the number of quantization bits for all layers and only calculated the output SNR as a measure of performance, as reported in the following subsection.  
	
	In our study, we have used MATLAB for the software simulation step of the workflow.

	\subsection{Implementation}
	We have considered three approached for the implementation: 1) hand-coded HDL, 2) Vivado HLS, and 3) a scalable RTL HDL code generator, as detailed below.
	
	\subsubsection{Hand-coded HDL}
	As a benchmark for comparison and a template for automatic and scalable code generation, the NN shown in Fig.~\ref{fig:NNarchitecture} was hand-coded in Verilog HDL. The register bit lengths were defined parametrically in the code, to facilitate the generalization of the HDL codes to arbitrary numbers of quantization bits.
	
	\subsubsection{High-level synthesis}
	In the next step, XILINX Vivado HLS tools were used to generate an HLS implementation of the NN. It is know that in Vivado, both C/C++ and SystemC are acceptable as input languages. Furthermore, Vivado HLS support fixed-point and floating-point arithmetic. Vivado generates RTL implementations of the modules in VHDL, Verilog, and SystemC \cite{nane2015survey}. Note that while native C-language data types are multiples of 8-bits (8, 16, 32, 64 bits), the RTL buses on the hardware can be of arbitrary lengths. For this, Vivado HLS tools allow the specification of arbitrary precision bit-width and do not rely on the native C data types. This option allows us to have a more accurate design in terms of power consumption and area \cite{VivadoDesignSuiteUserGuide}. Also, Vivado HLS tools allow RTL verification and C/RTL co-simulation in during the ''verification and fixed-point analysis'' phase. Using RTL verification and C/RTL co-simulation ensures that the HDL and HLS simulation are identical \cite{VivadoDesignSuiteUserGuide}.
	
	\subsubsection{HDL code generator}
	\label{sec:hdl_code_generator}
\begin{figure*}[tb]
		\centering\includegraphics[width=0.75\textwidth]{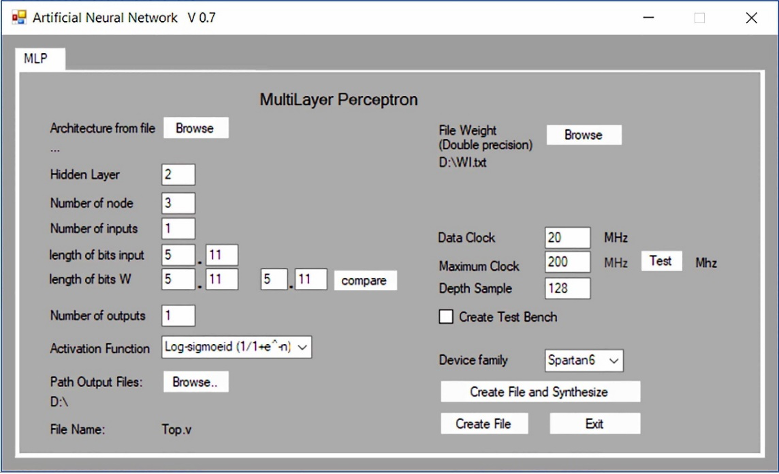}
		\caption{The GUI front-end of the C\# code generating software, which generates synthesizable Verilog HDL codes for neural networks of arbitrary depth, available online: \url{https://github.com/alphanumericslab/HDLCodeGenerators}}
		\label{fig:user_interface}
	\end{figure*}

\begin{figure*}[tb]
	\begin{subfigure}[a]{\linewidth}
		\centering\includegraphics[width=0.95\textwidth]{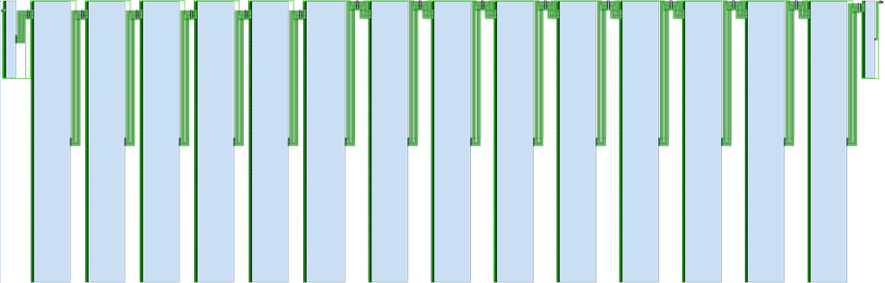}
    \caption{14 hidden-layer NN}
    \end{subfigure}
	\begin{subfigure}[b]{\linewidth}
		\centering\includegraphics[width=0.95\textwidth]{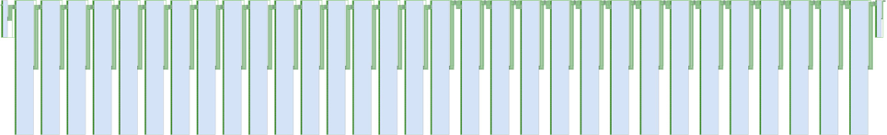}
    \caption{31 hidden-layer NN}
\end{subfigure}
		\caption{RTL schematics of NNs with 8 input and output nodes, with a) 14 and b) 31 32-node fully connected hidden layers, generated by the developed HDL NN generator}
		\label{fig:NNRTLSchematics}
	\end{figure*}

	A popular technique in software engineering is to develop high-level language scripts to generate codes in lower-level languages. This technique is highly advantageous for scalable algorithms, and repeated/parameterized code segments, for which human hand-coding is both time-taking and prone to errors. The repeated architecture of the NN is a good example, where the implementation of a single node or layer of the NN can be systematically replicated and interconnected with one another to obtain a complete NN architecture. For this, we have used the modules from our hand-coded implementations to develop a software in C\#, which gets the general specifications of the NN (the number of inputs, layers, outputs, node activation functions, NN weights file, quantization bits, etc.) through a user-interface and generates synthesizable HDL codes in Verilog corresponding to the required NN modules. The developed software is also able to compromise between resources and clock speed, depending on the selected system and data clock rates. The user-interface of the software is shown in Fig.~\ref{fig:user_interface}, and the major functions of this software, which generate the NN Verilog codes are listed in Table~\ref{tab:code_generator_main_functions}. In order to show the flexibility of the HDL code generator in producing NNs of arbitrary depth, Fig.~\ref{fig:NNRTLSchematics}, demonstrates the RTL schematic of two NNs with 8 inputs and outputs, and 14 and 31 fully connected hidden layers with 32 nodes per layer. The source codes of the developed software are online available in our GitHub repository: \href{https://github.com/alphanumericslab/HDLCodeGenerators}{https://github.com/alphanumericslab/HDLCodeGenerators}.
	
	\begin{table*}[tb]
		\caption{The major functions of the C\# software, which generate the NN architecture codes in Verilog}	
		\centering
		\begin{tabular}{|p{9.0cm}|p{7.5cm}|}\hline
			\texttt{Create\_TopModule(number\_of\_input, number\_of\_hidden\_Layer, number\_of\_node, number\_of\_output, length\_i, length\_w, clk\_max, clk\_data, path);} & Creates the top module of the NN, which can be instantiated in other modules or connected directly to the FPGA I/O busses\\\hline
			\texttt{Create\_Layer1(number\_of\_input, number\_of\_input, number\_of\_hidden\_Layer, number\_of\_mult, length\_i, length\_w, length\_o, path);} & Creates the input layer of the NN (between input and the first hidden layer)\\\hline
			\texttt{Create\_Layer(number\_of\_node,number\_of\_input, number\_of\_hidden\_Layer, number\_of\_mult, length\_i, length\_w, length\_o, path);} & Creates the intermediate hidden layers of the NN \\\hline
			\texttt{Create\_Layer\_End(number\_of\_node, number\_of\_output, number\_of\_hidden\_Layer, number\_of\_mult, length\_i, length\_w, length\_o, path);} & Creates the output layer of the NN (between last hidden layer and the output layer) \\\hline
			\texttt{Create\_AF(number\_of\_input,length\_i\_m+length\_w\_m, length\_i\_n + length\_w\_n,length\_i\_m, length\_i\_n, length\_w, path);} & Creates the codes for the activation function of the NN nodes \\\hline
			\texttt{Create\_AF\_End(number\_of\_output, length\_i\_m + length\_w\_m, length\_i\_n + length\_w\_n,length\_i\_m, length\_i\_n,length\_w, path);} & The activation function of the output layer (usually different from the intermediate layers of teh NN) \\\hline
			\texttt{Create\_mult(length\_i, length\_w, path);} & The multiply and adder units code generator \\\hline
		\end{tabular}
		\label{tab:code_generator_main_functions}
	\end{table*}
	
	\subsection{Verification and fixed-point analysis}
	The results of software simulations are compared with RTL and high-level implementations to verify the correctness of the hardware implementation by using arbitrary sets of weights and inputs, with different numbers of quantization bits.
	
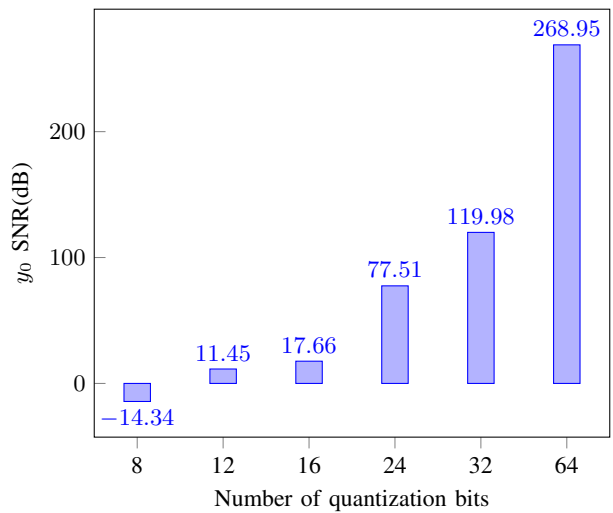
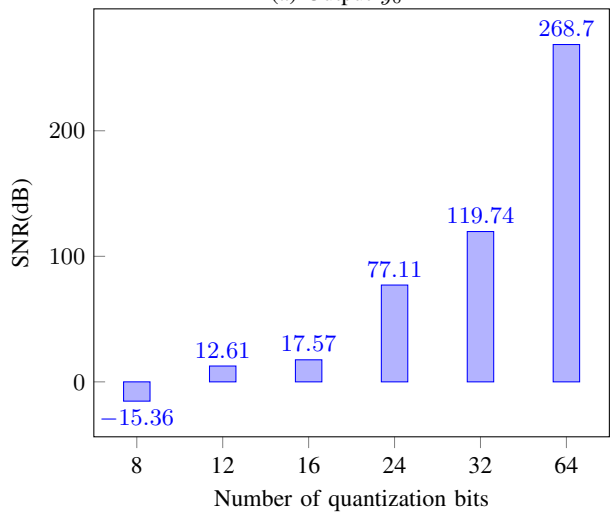
\begin{figure}[tbh]
		\begin{subfigure}{\linewidth}
		\small{
		\begin{tikzpicture}
			\begin{axis}[
			tick pos = left,
			ybar = 8pt,
			symbolic x coords={8,12,16,24,32,64},
			xlabel={Number of quantization bits},
			ylabel={$y_0$ SNR(dB)},
			nodes near coords,
			]
			\addplot coordinates {(8,-14.3378) (12,11.4541) (16,17.6619) (24,77.5099) (32,119.9818) (64,268.9472)};
			\end{axis}
			\end{tikzpicture}}
		\caption{Output $y_0$}
		\end{subfigure}	
		\begin{subfigure}{\linewidth}		\small{\begin{tikzpicture}
			\begin{axis}[
			tick pos = left,
			ybar = 8pt,
			symbolic x coords={8,12,16,24,32,64},
			xlabel={Number of quantization bits},
			ylabel={SNR(dB)},
			nodes near coords,
			]
			\addplot coordinates {(8,-15.3615) (12,12.6088) (16,17.5656) (24,77.1112) (32,119.7418) (64,268.7048)};
			\end{axis}
			\end{tikzpicture}}
			\caption{Output $y_1$}
		\end{subfigure}		
		\caption{The first ($y_0$) and second ($y_1$) NN outputs signal to noise ratios for different number of bits, as compared with the double-precision floating point simulation}
		\label{fig:y_SNR}
	\end{figure}

	For the studied NN architecture, the fixed-point analysis of the outputs ($y_0$, $y_1$) are shown in Fig.~\ref{fig:y_SNR}, where the fixed-point RTL implementations results are compared with double-precision simulations, in terms of SNR vs different fixed-point bit lengths. Accordingly, while the output SNR is negative with 8-bits of quantization (which is unacceptable, unless for trivial powers-of-two weights and inputs), the SNR improves as the number of quantization bits increase. As an extreme case, with 64 quantization bits, the HDL implementation approaches the double-precision floating-point accuracies. For the studied NN, depending on the application, fixed-point quantization bits between 20 and 24 are acceptable for most applications. 
	
	\subsection{Optimization}
	To this end, a synthesizable RTL/HSL implementation of the desired NN is obtained, which can be further optimized to improve timing parameters, power consumption, and area. A variety of optimization schemes can be used at this point, e.g., retiming, pipelining, repipelining, and C-slowing \cite{hauck2010reconfigurable}. As shown in Section~\ref{sec:state_space_optimization}, the state-space form permits the systematic implementation of these optimization schemes. The optimization can also be performed by adding synthesis tool optimization directives to the HDL, C/C++ or SystemC designs. Contemporary hardware design tools such as Vivado HLS, permit designers to add such directives to optimize the design by reducing latency and area and increasing parallelism \cite{koch2016fpgas}, \cite{VivadoHLSOptimizationMethodologyGuide}.
	\section{conclusion}
	In this work, a systematic workflow was proposed to convert iterative algorithms to hardware using state-space representations. It was shown how the state-space representation can be used for the efficient implementation and optimization of the resulting circuits, in combination with high-level synthesis tools.
	
	Shallow and deep neural networks were used as a case study for the proposed workflow. The case study was implemented using 1) hand-coded HDL, 2) XILINX Vivado high-level synthesis tools with synthesizable C/C++ and SystemC languages, and 3) a software that generates Verilog HDL codes of neural networks of arbitrary dimensions and depth  based on the parameters set by the user.

	The proposed scheme can be extended from various aspects. Considering the vast application of state-space formulations in signal processing and numerical analysis applications, the proposed workflow can be used to implement highly popular algorithms such as various filter architectures including the Kalman filter. The implementation of Long short-term memory (LSTM) recurrent neural networks, which can be formulated in terms of state-space models, is another emerging application.
	
	While the selection of the number of required quantization bits was performed by a statistical simulation with random inputs and NN weights, the procedure can be made more systematic by combining state-space models with analytical studies of quantization effects \cite{widrow2008quantization}.
	
	\bibliography{bib_ref}
	\bibliographystyle{IEEEtran}
	
\end{document}